\documentclass[conference]{IEEEtran}
\usepackage{cite}
\usepackage[pdftex]{graphicx}
\usepackage{amsmath}
\usepackage{amsfonts} %produces math-mode fonts
\usepackage{amsthm}
\usepackage{amssymb}
\usepackage[boxruled]{algorithm2e}

\usepackage{bbm}

\usepackage{kbordermatrix}

\newcommand{\pfun}{\mathop{\hbox{$\to$\kern-7pt\raise.9pt\hbox{\scalebox{1}[.55]{$|$}}\kern4pt} }}

\usepackage{array}

\begin{document}

\title{GraphChallenge.org \\ Triangle Counting Performance}

%\author{
%\IEEEauthorblockN{Hayden Jananthan}
%\IEEEauthorblockA{\textit{Department of Mathematics}\\
%\textit{Vanderbilt University}\\
%Nashville, Tennessee\\
%Email: hayden.r.jananthan@vanderbilt.edu}
%\and
%\IEEEauthorblockN{Jeremy Kepner}
%\IEEEauthorblockA{\textit{Lincoln Laboratory Supercomputing Center} \\
%\textit{Massachusetts Institute of Technology}\\
%Lexington, Massachusetts \\
%kepner@ll.mit.edu}
%%\and
%%\IEEEauthorblockN{Ziqi Zhou}
%%\IEEEauthorblockA{\textit{Mathematics Department}\\
%%\textit{Massachusetts Institute of Technology}\\
%%Cambridge, Massachusetts\\
%%zhouz@mit.edu}
%%\and
%%\IEEEauthorblockN{Suna Kim}
%%\IEEEauthorblockA{\textit{Mathematics Department} \\
%%\textit{California Institute of Technology}\\
%%Pasadena, California \\
%%skim3@caltech.edu}
%}
\author{\IEEEauthorblockN{Siddharth Samsi,
Jeremy Kepner, 
Vijay Gadepally,
Michael Hurley,
Michael Jones,
Edward Kao, \\
Sanjeev Mohindra, 
Albert Reuther,
Steven Smith,
William Song,
Diane Staheli,
Paul Monticciolo
\\
\IEEEauthorblockA{MIT Lincoln Laboratory, Lexington, MA}}}
\maketitle

\begin{abstract}
The rise of graph analytic systems has created a need for new ways to measure and compare
the capabilities of graph processing systems. The MIT/Amazon/IEEE Graph Challenge has been developed to provide a well-defined community venue for stimulating research and highlighting innovations in graph analysis software, hardware, algorithms, and systems.  GraphChallenge.org provides a wide range of pre-parsed graph data sets, graph generators, mathematically defined graph algorithms, example serial implementations in a variety of languages, and specific metrics for measuring performance.  The triangle counting component  of GraphChallenge.org tests the performance of graph processing systems to count all the triangles in a graph and exercises key graph operations found in many graph algorithms.  In 2017, 2018, and 2019 many triangle counting submissions were received from a wide range of authors and organizations.  This paper presents a performance analysis of the best performers of these submissions.   These submissions show that their state-of-the-art triangle counting  execution time, $T_{\rm tri}$, is a strong function of the number of edges in the graph, $N_e$, which improved significantly from 2017 ($T_{\rm tri} \approx (N_e/10^8)^{4/3}$) to 2018 ($T_{\rm tri} \approx N_e/10^9$) and remained comparable from 2018 to 2019.  Graph Challenge provides a clear picture of current graph analysis systems and underscores the need for new innovations to achieve high performance on very large graphs.
\end{abstract}

% \fi
\IEEEpeerreviewmaketitle

\section{Introduction}
\let\thefootnote\relax\footnotetext{This material is based upon work supported by the Assistant Secretary of Defense for Research and Engineering under Air Force Contract No. FA8702-15-D-0001 and National Science Foundation CCF-1533644. Any opinions, findings, conclusions or recommendations expressed in this material are those of the author(s) and do not necessarily reflect the views of the Assistant Secretary of Defense for Research and Engineering or the National Science Foundation.}

The importance of graph analysis has dramatically increased and is critical to a wide variety of domains that include the analysis of genomics \cite{Morrison2005,Mooney2012,polychronopoulos2014conserved,Dodson2014,Dodson2015,gouda2016distribution}, brain mapping \cite{fornito2016graph}, computer networks \cite{BrinPage1998,Faloutsos1999,yan2015spectrum,fontugne2017scaling,kepner2019new,kepner2019traffic}, social media \cite{Zuckerburg2005,Kwak2009}, cybersecurity \cite{shao2015percolation,yu2015malware}, and sparse machine learning \cite{lee2008sparse,boureau2008sparse,glorot2011deep,yu2012exploiting,Kepner2017graphblasDNN,kumar2018power9,alford2019sparseDNN}.

  Many graph processing systems are currently under development.  These systems are exploring innovations in algorithms \cite{Cormen2001,Miller2012,Buluc2014,voegele2017parallel,smith2017truss,hu2017trix,la2017ensemble,zhuzhunashvili2017preconditioned,low2017first,uppal2017scalable,mowlaei2017triangle,kepner2018mathematics}, software architecture \cite{BulucGilbert2011,Kepner2012-ch1,pearce2017triangle,halappanavar2017scalable,tom2017exploring,green2017quickly,kabir2017parallel,zhou2017design,hutchison2017distributed,wolf2017fast}, databases \cite{milechin2018d4mjl,cailliau2019redis}, software standards \cite{Mattson2013,kepner2015graphs,kepner2016mathematical,bulucc2017design,davis2017graphblas,kepner2019d4m,kepner2020graphblas}, and parallel computing hardware \cite{song2016novel,bisson2017static,date2017collaborative,debenedictis2017superstrider,manne2017if,kogge2017graph,gioiosa2017exploring,krawezik2018emu,kacher2020graphcore,lie2019cerebras}.
The rise of graph analysis systems has created a need for new ways to measure and compare
the capabilities of these systems. The MIT/Amazon/IEEE Graph Challenge has been developed to provide a well-defined community venue for stimulating research and highlighting innovations in graph analysis software, hardware, algorithms, and systems.   
GraphChallenge.org provides a wide range of pre-parsed graph data sets, graph generators, mathematically defined graph algorithms, example serial implementations in a variety of languages, and specific metrics for measuring performance.

Scale is an important driver of the Graph Challenge and graphs with billions to
trillions of edges are of keen interest.  The Graph Challenge is designed to work on
arbitrary graphs drawn from both real-world data sets and simulated data sets.
Examples of real-world data sets include the Stanford Large Network Dataset
Collection (see http://snap.stanford.edu/data), the AWS Public Data Sets (see
aws.amazon.com/public-data-sets), and the Yahoo! Webscope Datasets (see
webscope.sandbox.yahoo.com).  These real-world data sets cover a wide range of
applications and data sizes.  While real-world data sets have many contextual
benefits, synthetic data sets allow the largest possible graphs to be readily
generated. Examples of synthetic data sets include Graph500, Block Two-level
Erdos-Renyi graph model (BTER) \cite{seshadhri2012community}, Kronecker Graphs
\cite{kepner2011graph,sanders2018triangle,kepner2019radixnet}, and  Perfect Power Law graphs
\cite{kepner2012perfect,gadepally2015using,kepner2018powerlaw}. The focus of the Graph Challenge is on graph analytics.  While parsing and formatting
complex graph data are necessary in any graph analysis system, these data sets are
made available to the community in a variety of pre-parsed formats to minimize the
amount of parsing and formatting  required by Graph Challenge participants.  The public data
are available in a variety of formats, such as linked list, tab separated, and
labeled/unlabeled.

Graph Challenge 2017 received a large number of submissions that highlighted innovations in hardware, software, algorithms, systems, and visualization that allows the state-of-the-art in graph processing for 2017 to be estimated \cite{GraphChallenge2017analysis}.  The goal of this paper is to analyze and synthesize the 2018 and 2019 submissions to provide an updated picture of the current state of the art of graph analysis systems.  The organization of this paper is as follow.  First, a recap of triangle counting is provided, along with a few standard algorithms.  Next, an overview is presented of the Graph Challenge 2018 and 2019 submissions.  The core of the paper is the section on the analysis of the 17 submission that all performed the triangle counting challenge.  Based on this analysis, these results are synthesized to provide a picture of the current state of the art.

\section{Triangle Counting}

The Graph Challenge consists of three challenges
\begin{itemize}
\item Pre-challenge: PageRank pipeline \cite{dreher2016pagerank}
\item Static graph challenge: subgraph isomorphism \cite{samsi2017static}
\item Streaming graph challenge: stochastic block partition \cite{ed}
\item Sparse deep neural network challenge \cite{SparseDNNchallenge}
\end{itemize}
The static graph challenge is further broken down into triangle counting and k-truss.  Triangle counting is the focus of this paper.

\begin{figure}[ht]
\centering
\includegraphics[width=\columnwidth]{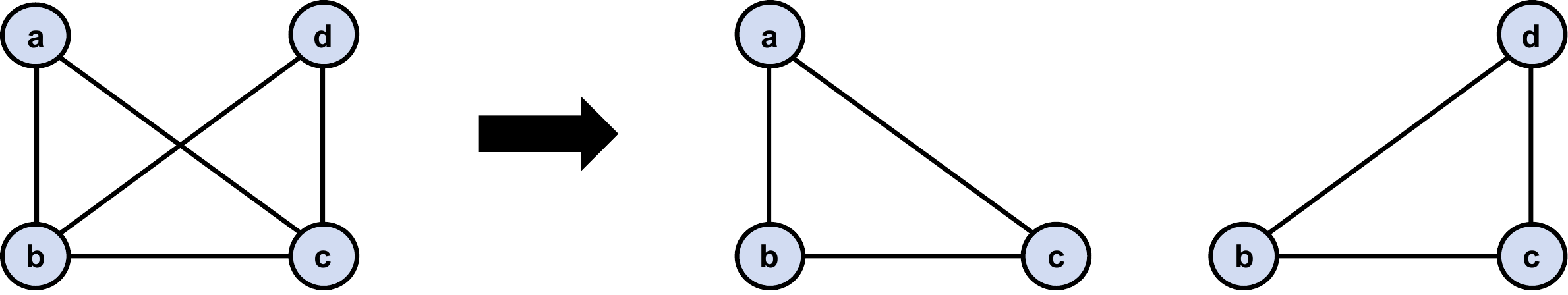}
\caption{The graph shown in this example contains two triangles consisting of nodes \{a,b,c\} and \{b,c,d\}.}
\label{fig:triangle}
\end{figure}

Triangles are the most basic, trivial sub-graph. A triangle can be defined as a
set of three mutually adjacent vertices in a graph. As shown in Figure~\ref{fig:triangle}, the graph \textbf{G} contains two triangles comprising nodes \{a,b,c\} and \{b,c,d\}. The number of triangles in a graph
is an important metric used in applications such as social network mining, link
classification and recommendation, cyber security, functional biology, and spam
detection ~\cite{pavan2013}.

The number of triangles in a given graph \textbf{G} can be calculated in several
ways. We highlight two algorithms based on linear algebra primitives. The first
algorithm proposed by Wolf et al~\cite{wolf2015} uses an overloaded matrix
multiplication approach on the adjacency and incidence matrices of the graph and is
shown in Algorithm~\ref{triangle_1}. The second approach proposed by Burkhardt
et al~\cite{burkhardt2016} uses only the adjacency matrix of the given graph and is
shown in Algorithm~\ref{triangle_2}.

\begin{algorithm}[h]
  \KwData{Adjacency matrix \textbf{A} and incidence matrix \textbf{E}}
  \KwResult{Number of triangles in graph \textbf{G}}
  initialization\;
  $\textbf{C} = \textbf{A}\textbf{E}$\\
  $n_{T} = nnz(C)/3$\\
  \vspace{.25cm}
  Multiplication is overloaded such that \\
  $\textbf{C}(i,j) = \{i, x, y\}$ iff \\
  $\textbf{A}(i,x) = \textbf{A}(i,y) = 1$ \& $\textbf{E}(x,j) = \textbf{E}(y,j) = 1$ \\
  \vspace{.2cm}
  \caption{Array based implementation of triangle counting algorithm using the adjacency and incidence matrix of a graph~\cite{wolf2015}.}
  \label{triangle_1}
\end{algorithm}

\begin{algorithm}[ht]
\KwData{Adjacency matrix \textbf{A}}
\KwResult{Number of triangles in graph \textbf{G}}
initialization\;
$\textbf{C} = \textbf{A}^2 \circ \textbf{A}$ \\
$n_{T} = \sum_{ij}^{ } (\textbf{C}) / 6$ \\
\vspace{1em}
Here, $\circ$ denotes element-wise multiplication\\
\vspace{.2cm}
\caption{Array based implementation of triangle counting algorithm using only the adjacency matrix of a graph~\cite{burkhardt2016}.}
\label{triangle_2}
\end{algorithm}

Another algorithm for triangle counting based on a masked matrix multiplication
approach has been proposed by Azad et al~\cite{gilbert2015}. The serial version of
this algorithm based on the MapReduce implementation by Cohen et al~\cite{cohen2009}
is shown in Algorithm~\ref{triangle_3}. Finally, a comparison of triangle counting
algorithms can be found in~\cite{wang2016}.

\begin{algorithm}[h]
\KwData{Adjacency matrix \textbf{A}}
\KwResult{Number of triangles in graph \textbf{G}}
initialization\;
$(\textbf{L}, \textbf{U}) \leftarrow \textbf{A}$\\
$\textbf{B} = \textbf{L}\textbf{U} $ \\
$\textbf{C} = \textbf{A} \circ \textbf{B} $ \\
$n_{T} = \sum_{ij}^{ } (\textbf{C}) / 2$ \\
\vspace{1em}
Here, $\circ$ denotes element-wise multiplication\\
\vspace{.2cm}
\caption{Serial version of triangle counting algorithm based on MapReduce version by Cohen et al~\cite{cohen2009} and ~\cite{gilbert2015}.}
\label{triangle_3}
\end{algorithm}

\section{Community Submissions}

Graph Challenge has received a wide range of submissions across all its various challenges that have included hundreds of authors from over fifty organizations.  In 2018, eighteen submissions were selected for publication \cite{Bisson-Nvidia-2018,Hu-GWU-2018,Yasar-GaTech-2018,Fox-LLNL-2018,Mailthody-UIUC-2018,Zhang-CMU-2018,Davis-TAMU-2018,Donato-UMassB-2018,Kuo-CUHK-2018,Pearce-LLNL-2018,Conte-NII-2018,Low-CMU-2018,Fox-LLNL-2018b,Green-GaTech-2018,Ghosh-WashSt-2018,Huang-UIUC-2018,Sadi-CMU-2018,Uppal-GWU-2018} and nine provided sufficient triangle counting performance data for analysis \cite{Bisson-Nvidia-2018,Hu-GWU-2018,Yasar-GaTech-2018,Fox-LLNL-2018,Mailthody-UIUC-2018,Zhang-CMU-2018,Davis-TAMU-2018,Donato-UMassB-2018,Kuo-CUHK-2018}.  In 2019, twenty submissions were selected for publication \cite{Pandey-Stevens-2019,Pearce-LLNL-2019,Acer-Sandia-2019,Yasar-GaTech-2019,Hoang-UTexas-2019,Wang-UCDavis-2019,Gui-HuazhongU-2019,Pearson-UIUC-2019,Bisson-Nvidia-2019,Blanco-CMU-2019,Davis-TAMU-2019,Ellis-Sandia-2019,Ghosh-PNNL-2019,Liu-PNNL-2019,Almasri-UIUC-2019,Wang-UCDavis-2019b,Wanye-VaTech-2019,Wang-PingAn-2019,Huang-UIUC-2019,Mofrad-UPitt-2019} and eight provided sufficient triangle counting performance data for analysis \cite{Pandey-Stevens-2019,Pearce-LLNL-2019,Acer-Sandia-2019,Yasar-GaTech-2019,Hoang-UTexas-2019,Wang-UCDavis-2019,Gui-HuazhongU-2019,Pearson-UIUC-2019}.

Numerous submissions implemented the triangle counting challenge in a comparable manner, resulting in over 800 distinct measurements of triangle counting execution time, $T_{\rm tri}$.  The number of edges, $N_e$, in the graph describes the overall size of the graph.  The rate of edges processed in triangle counting is given by
$$
 {\rm Rate} = N_e/T_{\rm tri}
$$

Analyzing and combining all the performance data from the submissions can be done by fitting a piecewise model to each submission and then comparing the models.  For each submission, $T_{\rm tri}$ vs $N_e$ is plotted on a log-log scale from which a  model can be fit to the data by estimating the parameters $N_1$ and $\beta$ in the formula
$$
   T_{\rm tri} = (N_e/N_1)^\beta
$$
where $N_1$ is the number edges that can be processed in 1 second. The triangle counting execution time vs number of edges and corresponding model fits are given in Appendix A. The model fits illustrate the strong dependence of $T_{\rm tri}$ on $N_e$.   

\section{Performance Analysis}

\begin{figure}[ht]
\centering
\includegraphics[width=\columnwidth]{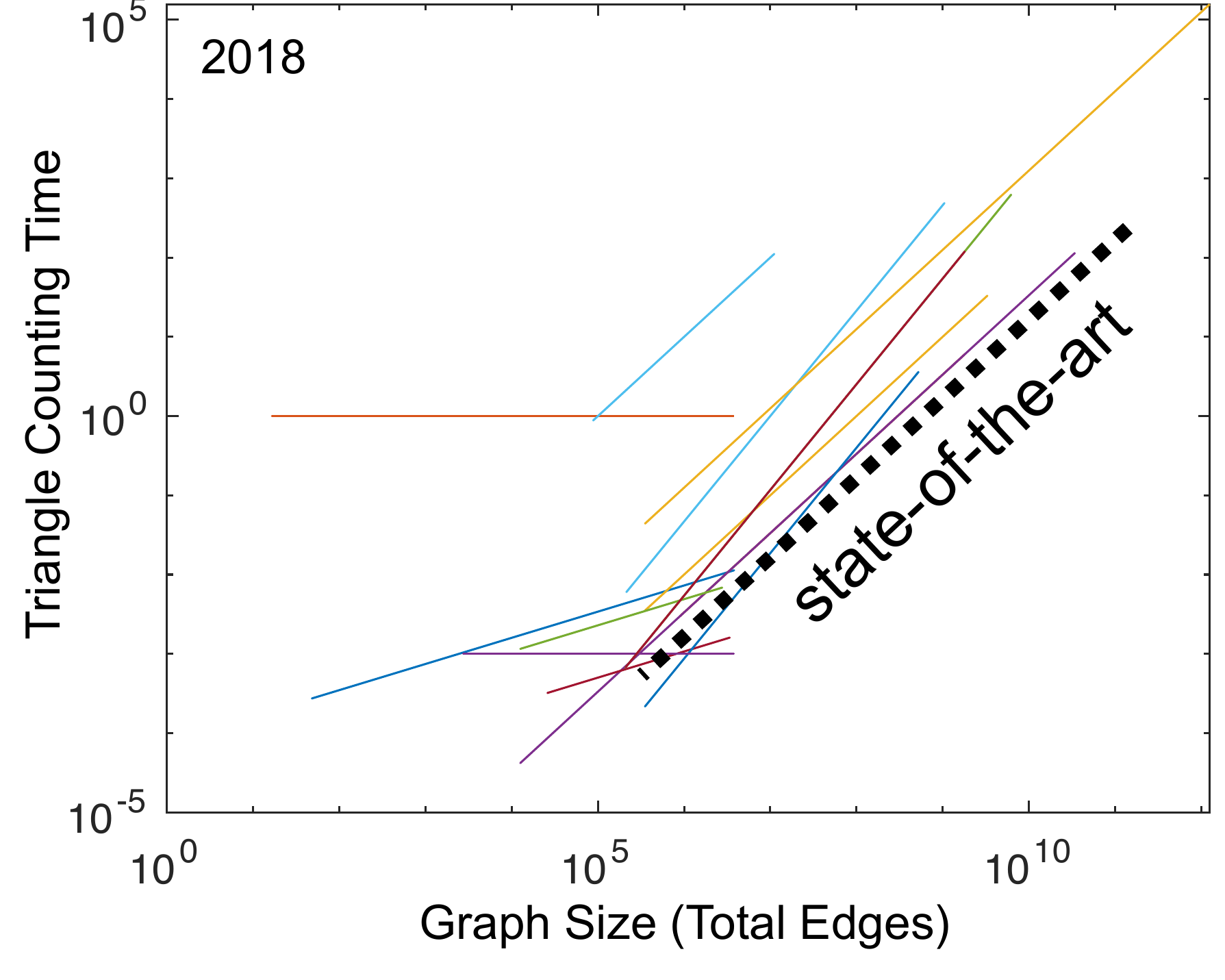}
\includegraphics[width=\columnwidth]{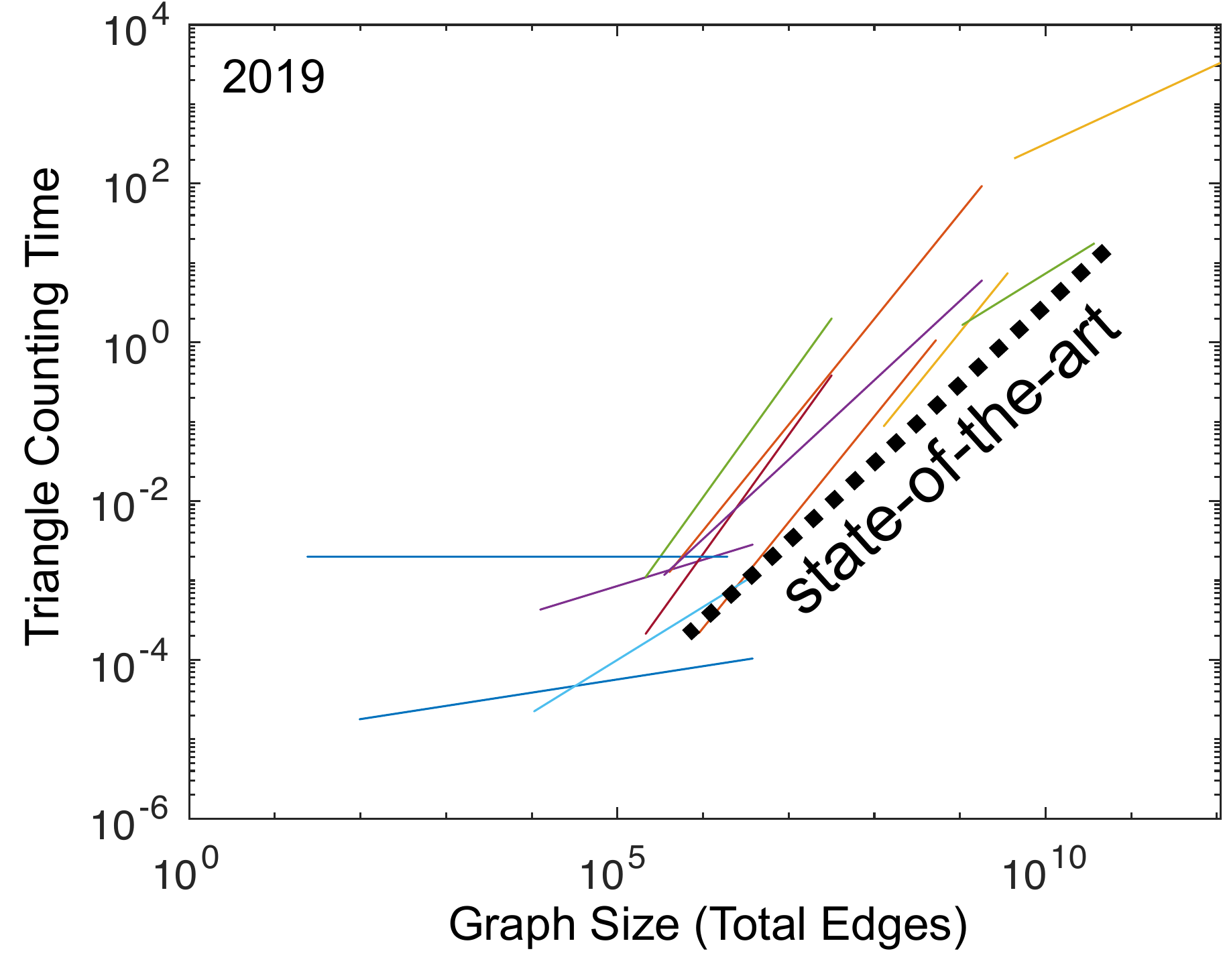}
\caption{Model fits of triangle execution time vs number edges for selected Graph Challenge 2018 (top) and 2019 (bottom) triangle counting submissions.  State-of-the-art is denoted by the black dashed line.}
\label{fig:TimePerformance}
\end{figure}

\begin{figure}[ht]
\centering
\includegraphics[width=\columnwidth]{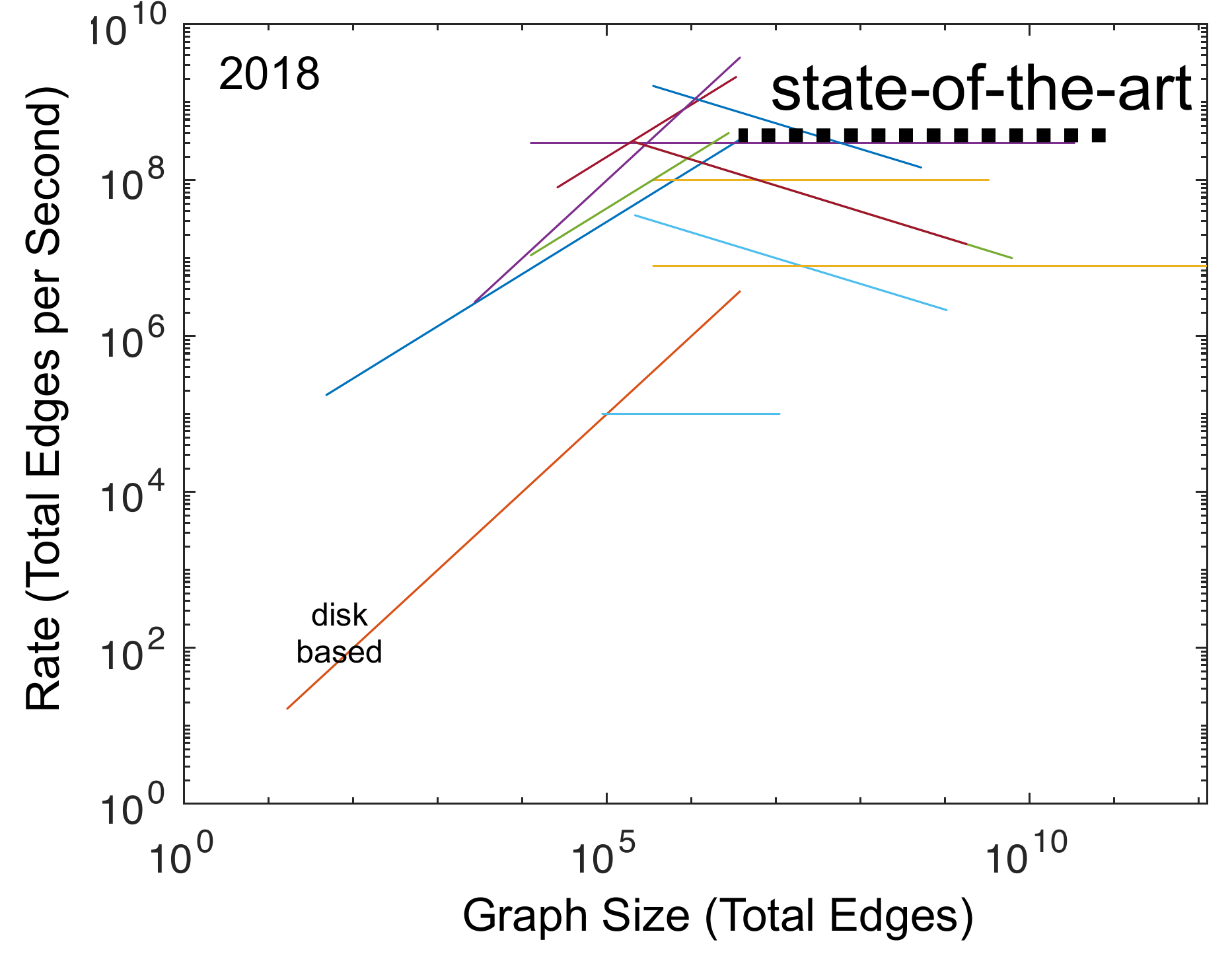}
\includegraphics[width=\columnwidth]{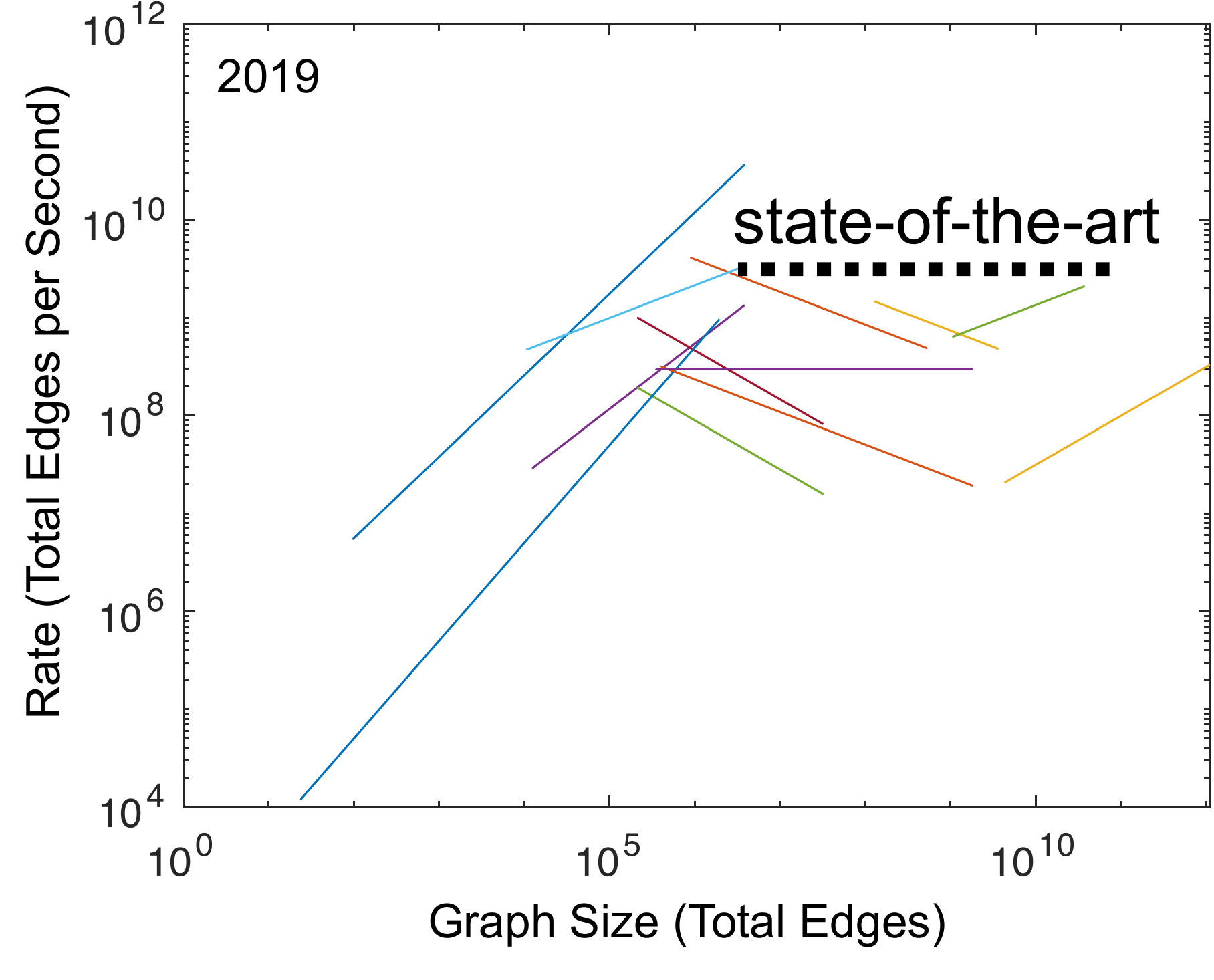}
\caption{Model fits of triangle execution rate vs. number edges for selected Graph Challenge 2018 (top) and 2019 (bottom) triangle counting submissions.  State-of-the-art is denoted by the black dashed line.}
\label{fig:RatePerformance}
\end{figure}

The normalized parameters $N_1$ and $\beta$, along with the largest values of $N_e$, are shown in  Tables~\ref{table:NormalizeModel2018} and \ref{table:NormalizeModel2018} for each submission.  Submissions with larger $N_e$, larger $N_1$, and smaller $\beta$ perform best.  The current state-of-the-art can be seen by plotting all the model fits $T_{\rm tri}$ together (see Figures~\ref{fig:TimePerformance} and \ref{fig:RatePerformance}).   Combined, these suggest that state-of-the-art performance model of the 2018 and 2019 is
$$
   T_{\rm tri} \approx N_e/10^9
$$
which is a significant improvement over the 2017 state-of-the-art performance model of
$$
   T_{\rm tri} \approx (N_e/10^8)^{4/3}
$$
Given the enormous diversity in processors, algorithms, and software, this relatively consistent picture of the state-of-the-art suggests that the current limitations are set by common elements across these benchmarks, such as memory bandwidth.

\begin{table}
\caption{{\rm 2018 Triangle counting time model fit coefficients for $T_{\rm tri} = (N_e/N_1)^\beta$ for large values of  $N_e$.}}
\centering
\begin{tabular}{lllcc}
\hline
Ref & Submission & max $N_e$ & $N_1$ & $\beta$ \\
\hline
\cite{Bisson-Nvidia-2018}          & Bisson-Nvidia-2018       & $1.8\times10^9$    & $3\times10^8$ & $1$ \\
\cite{Hu-GWU-2018}                 & Hu-GWU-2018              & $3.4\times10^{10}$ & $3\times10^8$ & $1$ \\
\cite{Yasar-GaTech-2018}           & Yasar-GaTech-2018        & $3.3\times10^{9}$  & $1\times10^8$ & $1$ \\

\cite{Fox-LLNL-2018}               & Fox-LLNL-2018            & $5.2\times10^{8}$  & $2\times10^8$ & $4/3$ \\
\cite{Mailthody-UIUC-2018}         & Mailthody-UIUC-2018      & $1.0\times10^{9}$  & $1\times10^7$ & $4/3$ \\
\cite{Zhang-CMU-2018}              & Zhang-CMU-2018           & $3.4\times10^{10}$ & $8\times10^6$ & $1$ \\

\cite{Davis-TAMU-2018}             & Davis-TAMU-2018          & $1.8\times10^{9}$ & $5\times10^7$ & $4/3$ \\
\cite{Donato-UMassB-2018}          & Donato-UMassB-2018       & $6.2\times10^{9}$ & $5\times10^7$ & $4/3$ \\
\cite{Kuo-CUHK-2018}               & Kuo-CUHK-2018            & $1.1\times10^{7}$ & $1\times10^5$ & $1$ \\
\hline
\end{tabular}
\label{table:NormalizeModel2018}
\end{table}

\begin{table}
\caption{{\rm 2019 Triangle counting time model fit coefficients for $T_{\rm tri} = (N_e/N_1)^\beta$ for large values of  $N_e$.}}
\centering
\begin{tabular}{lllcc}
\hline
Ref & Submission & max $N_e$ & $N_1$ & $\beta$ \\
\hline
\cite{Pandey-Stevens-2019}         & Pandey-Stevens-2019      & $5.2\times10^8$    & $5\times10^8$ & $4/3$ \\
\cite{Pearce-LLNL-2019}            & Pearce-LLNL-2019         & $1.1\times10^{12}$ & $5\times10^5$ & $1/2$ \\
\cite{Acer-Sandia-2019}            & Acer-Sandia-2019         & $3.6\times10^{9}$  & $6\times10^7$ & $3/2$ \\
\cite{Yasar-GaTech-2019}           & Yasar-GaTech-2019        & $1.8\times10^{9}$  & $3\times10^8$ & $1$ \\

\cite{Hoang-UTexas-2019}           & Hoang-UTexas-2019        & $3.7\times10^{10}$ & $5\times10^8$ & $2/3$ \\
\cite{Wang-UCDavis-2019}           & Wang-UCDavis-2019        & $3.2\times10^{7}$  & $2\times10^7$ & $3/2$ \\
\cite{Gui-HuazhongU-2019}          & Gui-HuazhongU-2019       & $3.2\times10^{7}$  & $6\times10^7$ & $3/2$ \\
\cite{Pearson-UIUC-2019}           & Pearson-UIUC-2019        & $1.8\times10^{9}$  & $6\times10^7$ & $4/3$ \\
\hline
\end{tabular}
\label{table:NormalizeModel2019}
\end{table}

\section{Conclusion}

The rapid increase in the use of graphs and has inspired new ways to measure and compare
the attributes of graph analytic systems. The MIT/Amazon/IEEE Graph Challenge was created to stimulate research  in graph analysis software, hardware, algorithms, and systems.  The GraphChallenge.org website makes available to the world many pre-processed graph data sets, graph generators, graph algorithms, prototype serial implementations in a several languages, and defined metrics for assessing performance.  The triangle counting component  of GraphChallenge.org tests the performance of graph processing systems to count all the triangles in a graph and exercises key graph operations found in many graph algorithms.  In 2017, 2018, and 2019 many triangle counting submissions were received from a wide range of authors and organizations.   These submissions show that their state-of-the-art triangle counting  execution time, $T_{\rm tri}$, is a strong function of the number of edges in the graph, $N_e$, which improved significantly from 2017 ($T_{\rm tri} \approx (N_e/10^8)^{4/3}$) to 2018 ($T_{\rm tri} \approx N_e/10^9$) and remained comparable from 2018 to 2019.  Graph Challenge provides a clear picture of current graph analysis systems and underscores the need for new innovations to achieve high performance on very large graphs.


\section*{Acknowledgments}

The authors wish to acknowledge the following individuals for their contributions and support: Alan Edelman, Charles Leiserson, Steve Pritchard, Michael Wright, Bob Bond, Dave Martinez, Sterling Foster, Paul Burkhardt,  Victor Roytburd, Trung Tran, along with William Arcand, David Bestor, William Bergeron, Chansup Byun, Matthew Hubbell, Michael Houle, Anna Klein, Peter Michaleas, Lauren Milechin, Julie Mullen, Andrew Prout, Antonio Rosa, and Charles Yee.

% The ''triggered'' command can be changed if desired:


%\bibliographystyle{IEEEtran}
\bibliographystyle{ieeetr}
\bibliography{aarabib}
%\begin{thebibliography}{1}

\appendices
\setcounter{equation}{0}
\renewcommand{\theequation}{\thesection\arabic{equation}}

\section{2018 Triangle Counting Submissions}
\label{sec:2018submissions}

\begin{figure}[ht]
\centering
\includegraphics[width=2.5in]{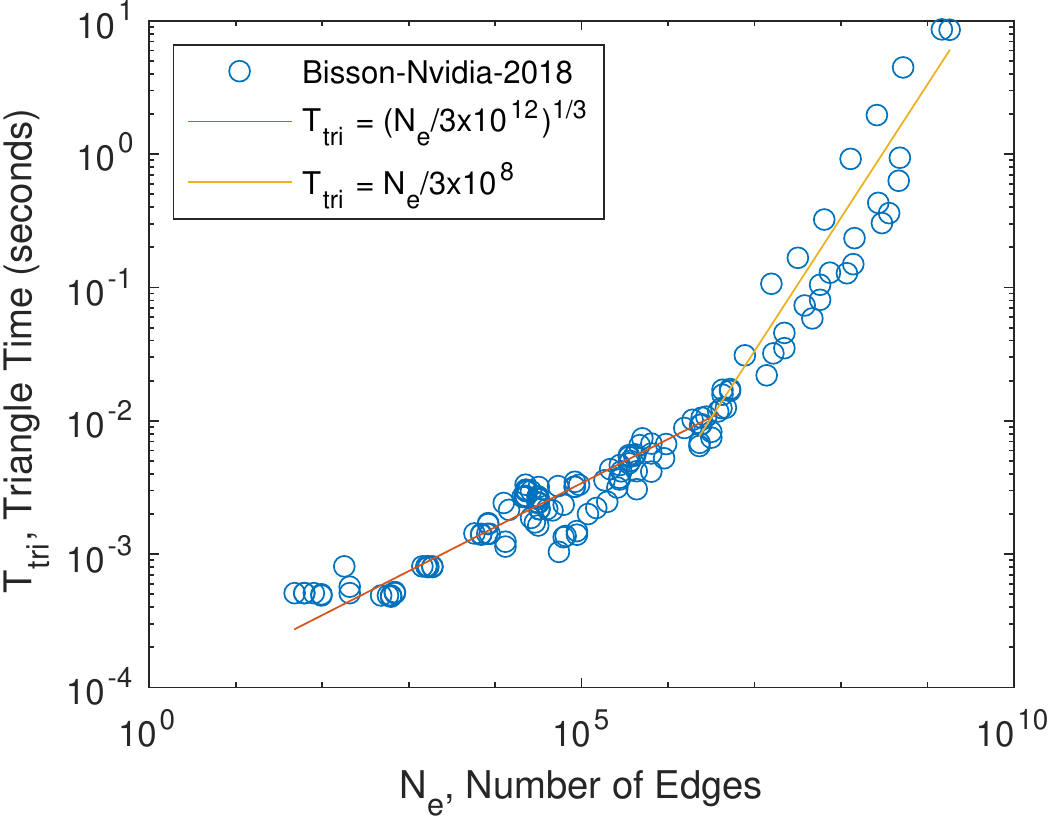}
\includegraphics[width=2.5in]{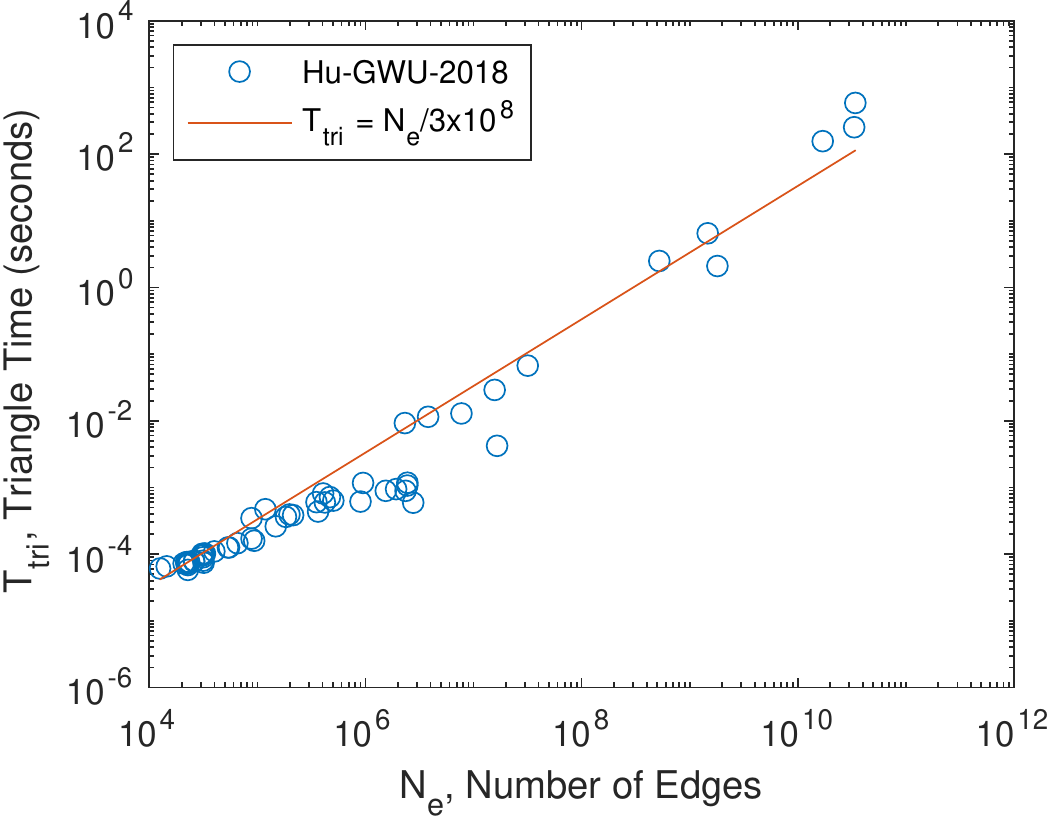}
\includegraphics[width=2.5in]{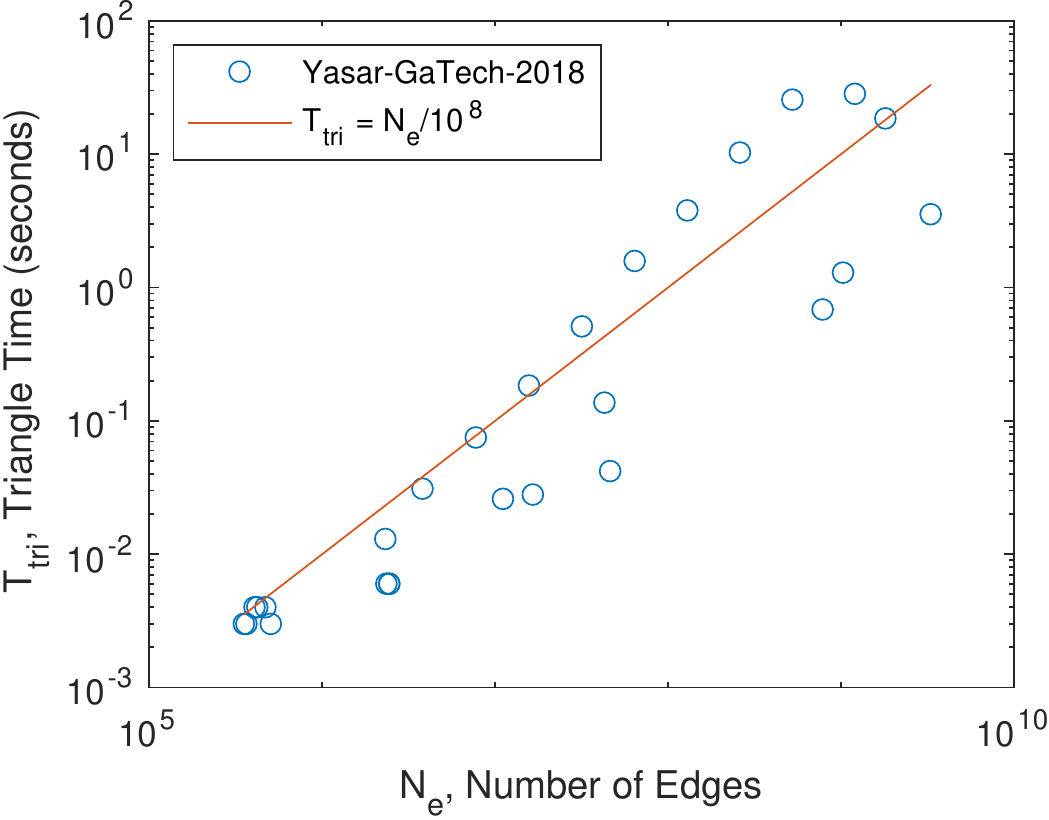}
\caption{Graph Challenge 2018 Champions. Triangle counting execution time vs number of edges and corresponding model fits for Bisson-Nvidia-2018 \cite{Bisson-Nvidia-2018}, Hu-GWU-2018 \cite{Hu-GWU-2018}, and Yasar-GaTech-2018 \cite{Yasar-GaTech-2018}.}
\label{fig:Champions}
\end{figure}

\begin{figure}[ht]
\centering
\includegraphics[width=2.5in]{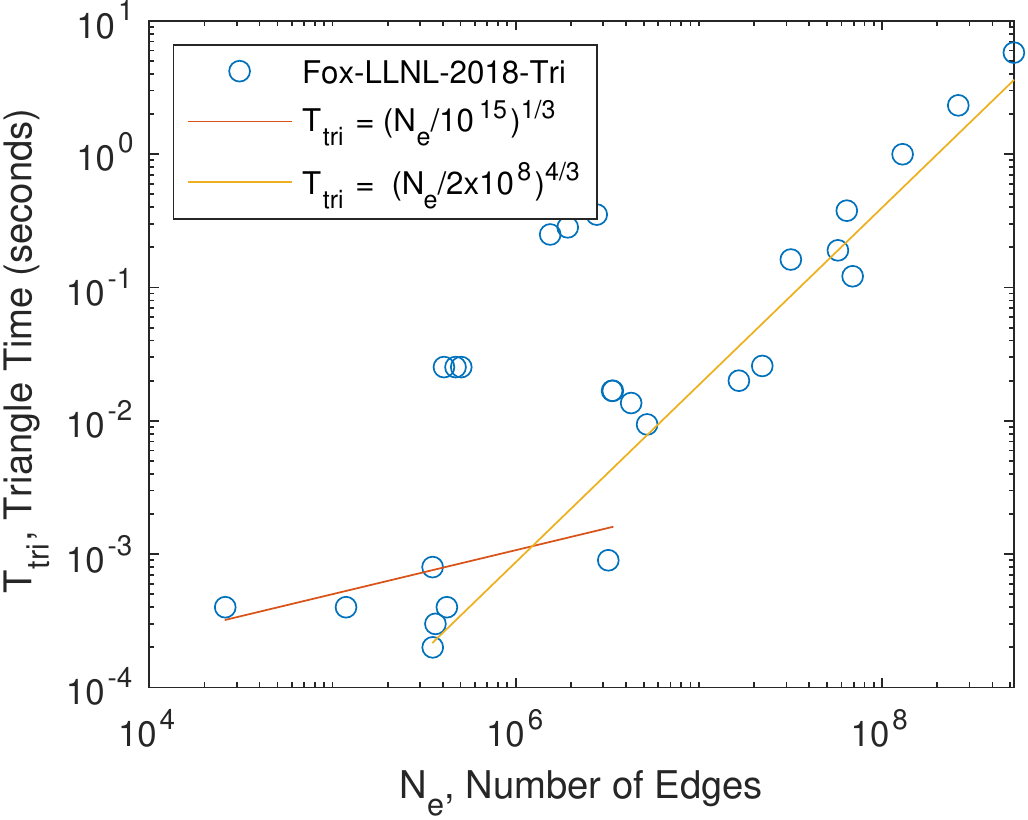}
\includegraphics[width=2.5in]{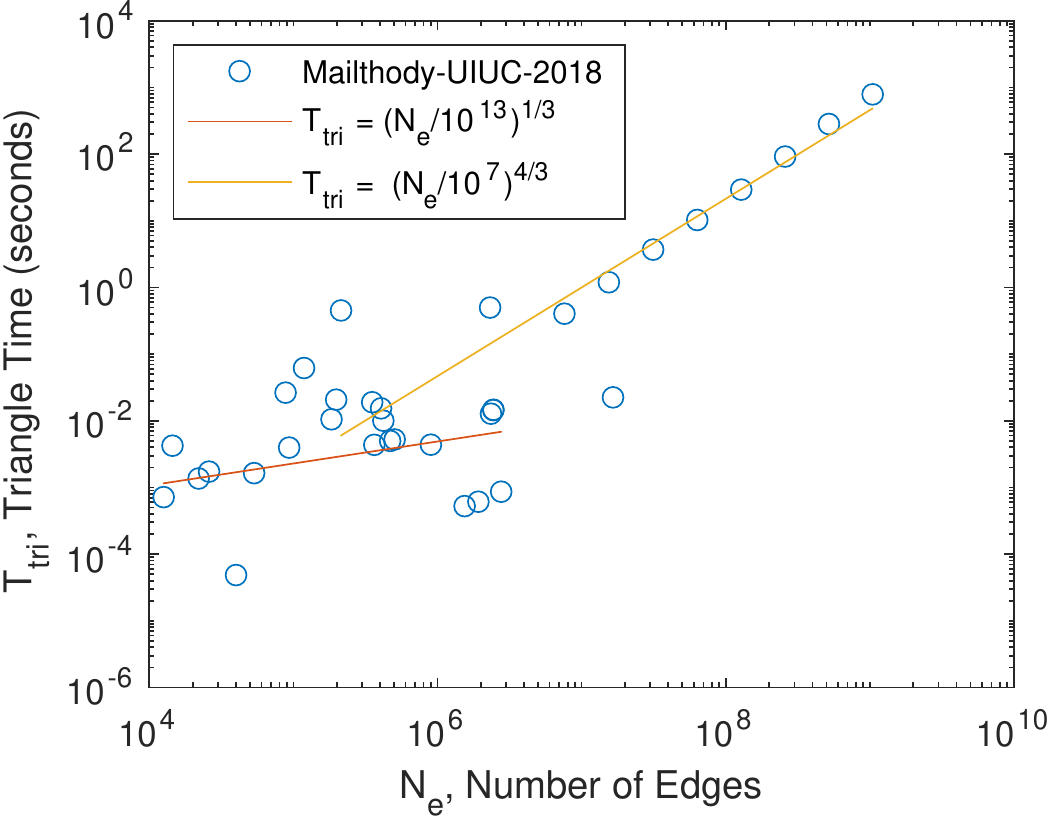}
\includegraphics[width=2.5in]{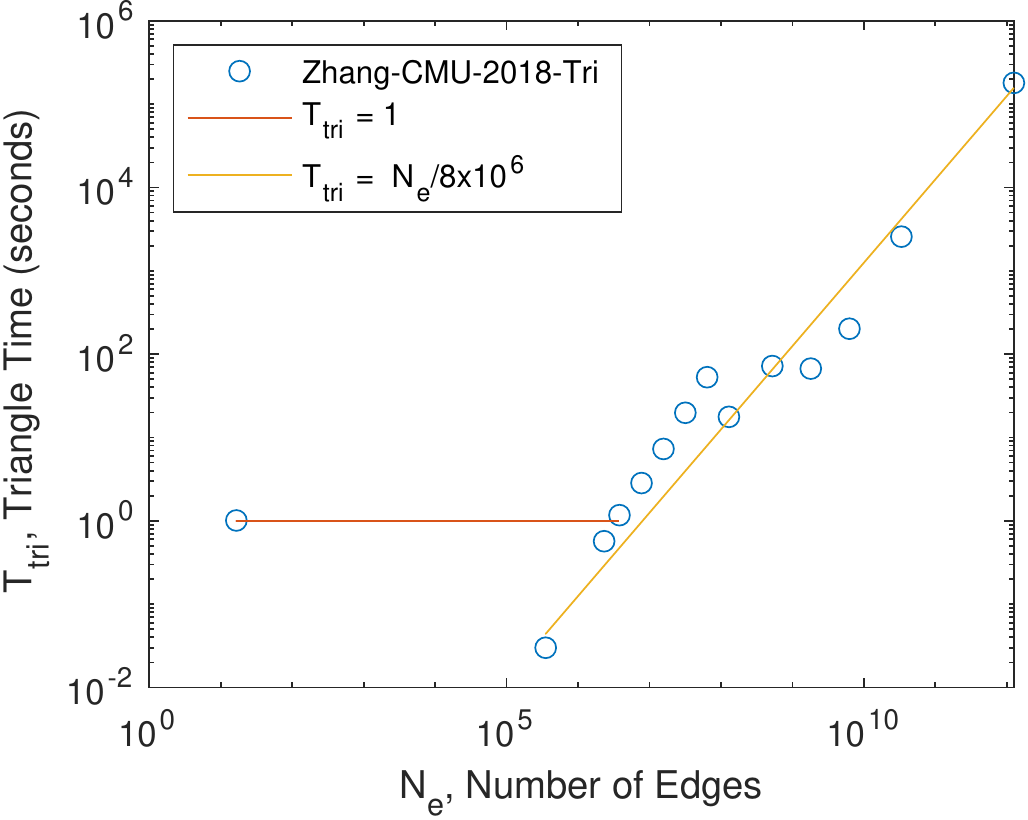}
\caption{Graph Challenge 2018 Finalists. Triangle counting execution time vs number of edges and corresponding model fits for Fox-LLNL-2018 \cite{Fox-LLNL-2018}, Mailthody-UIUC-2018 \cite{Mailthody-UIUC-2018}, and Zhang-CMU-2018 \cite{Zhang-CMU-2018}.}
\label{fig:Finalists}
\end{figure}

\begin{figure}[ht]
\centering
\includegraphics[width=2.5in]{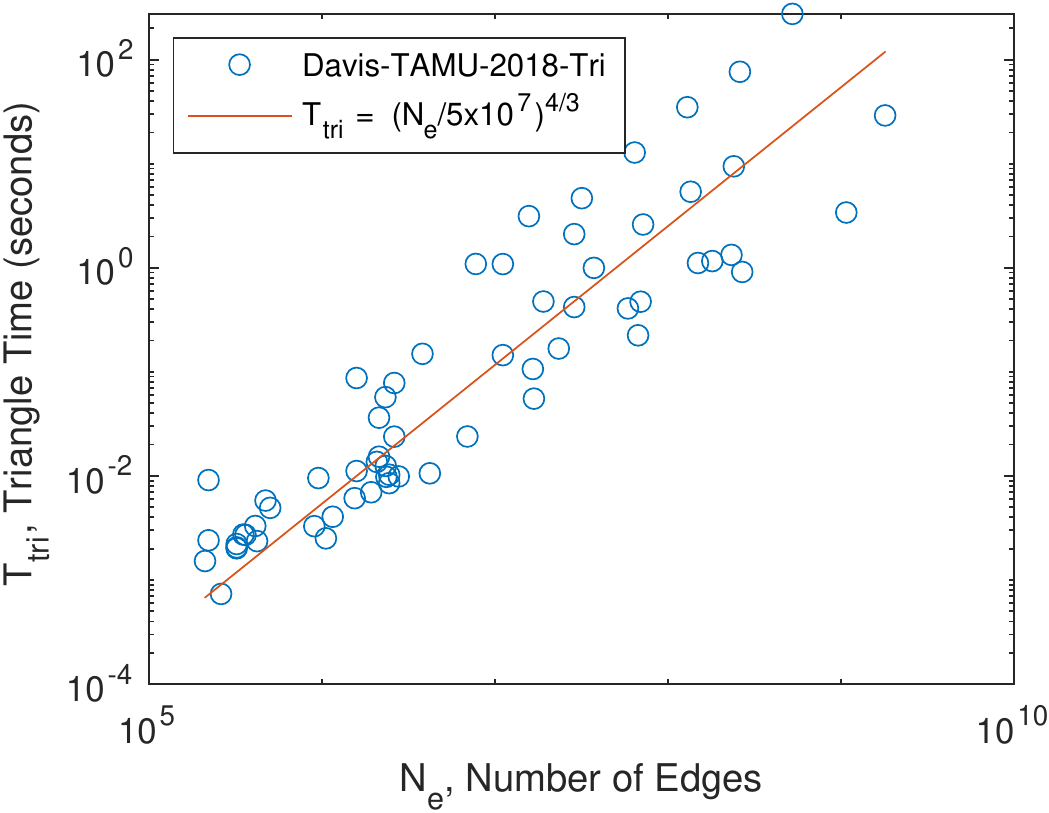}
\includegraphics[width=2.5in]{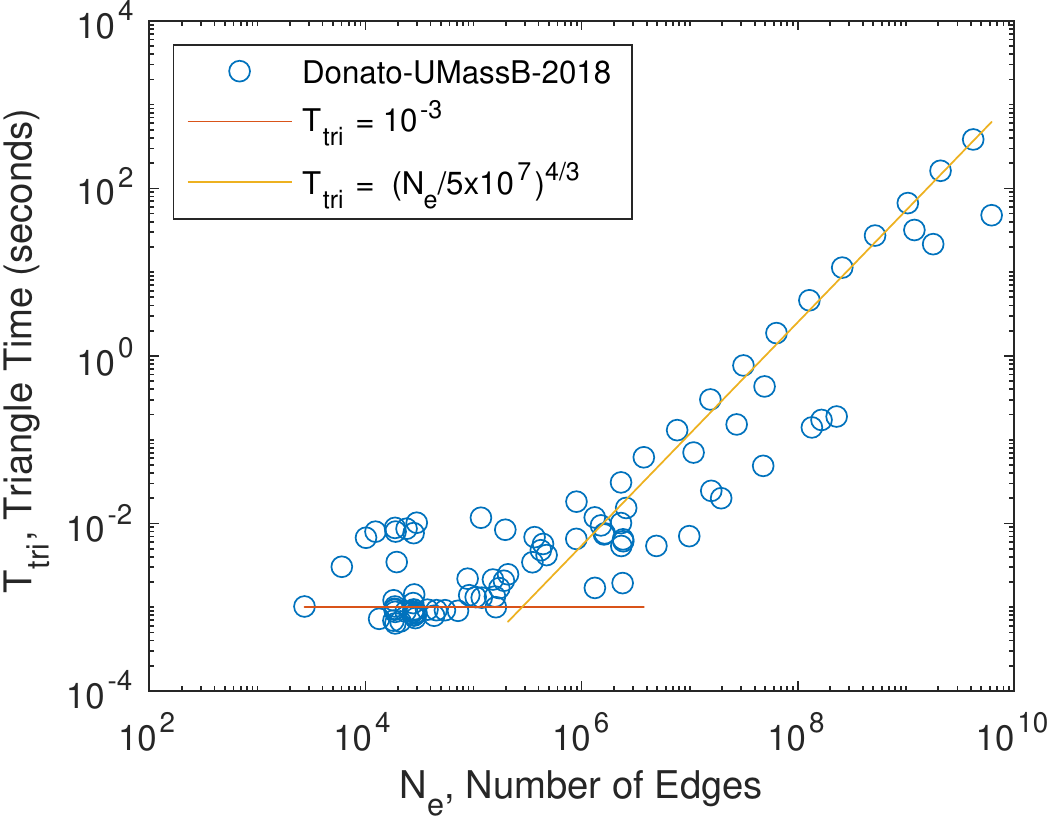}
\includegraphics[width=2.5in]{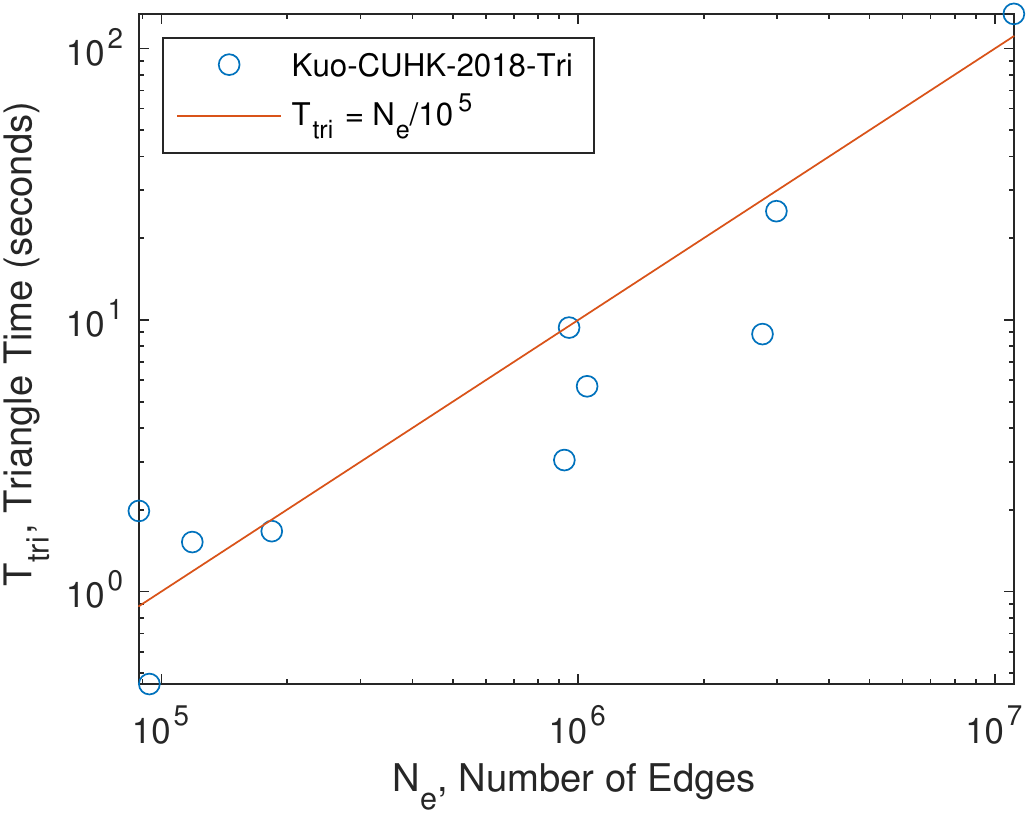}
\caption{Graph Challenge 2018 Innovation Award and Honorable Mentions. Triangle counting execution time vs number of edges and corresponding model fits for Davis-TAMU-2018 \cite{Davis-TAMU-2018}, Donato-UMassB-2018 \cite{Donato-UMassB-2018}, and Kuo-CUHK-2018 \cite{Kuo-CUHK-2018}.}
\label{fig:Finalists}
\end{figure}

\clearpage

\section{2019 Triangle Counting Submissions}
\label{sec:2019submissions}

\begin{figure}[ht]
\centering
\includegraphics[width=2.4in]{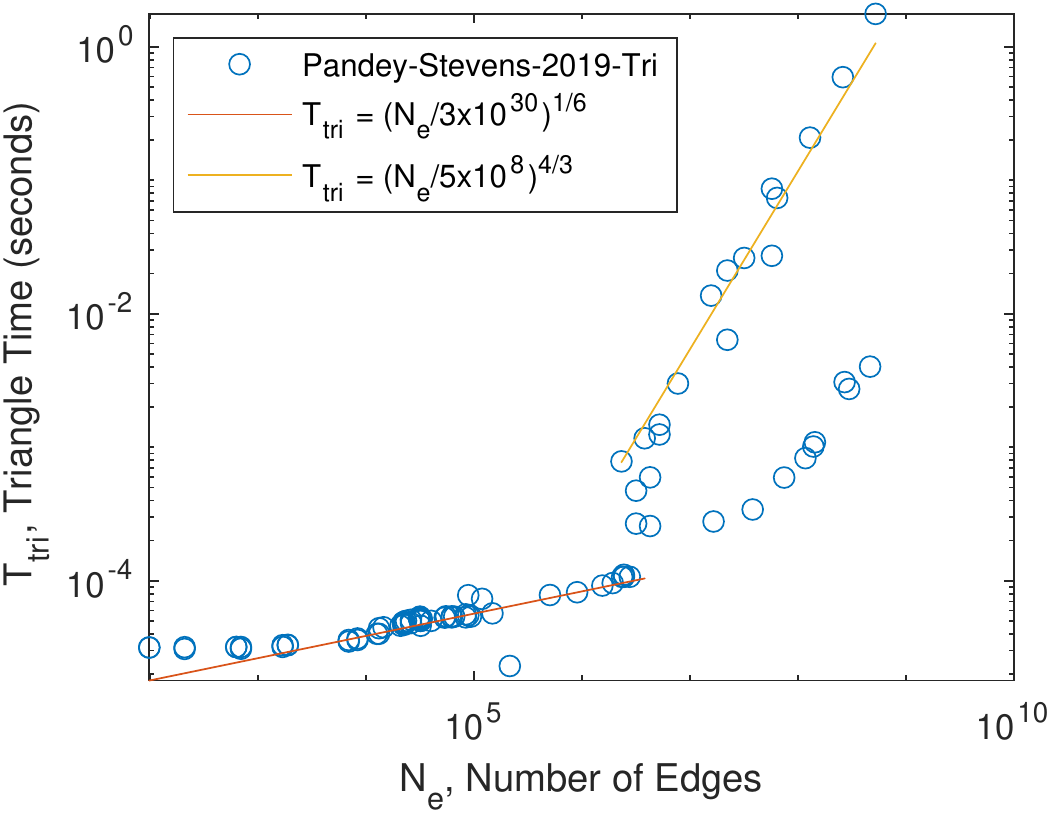}
\includegraphics[width=2.4in]{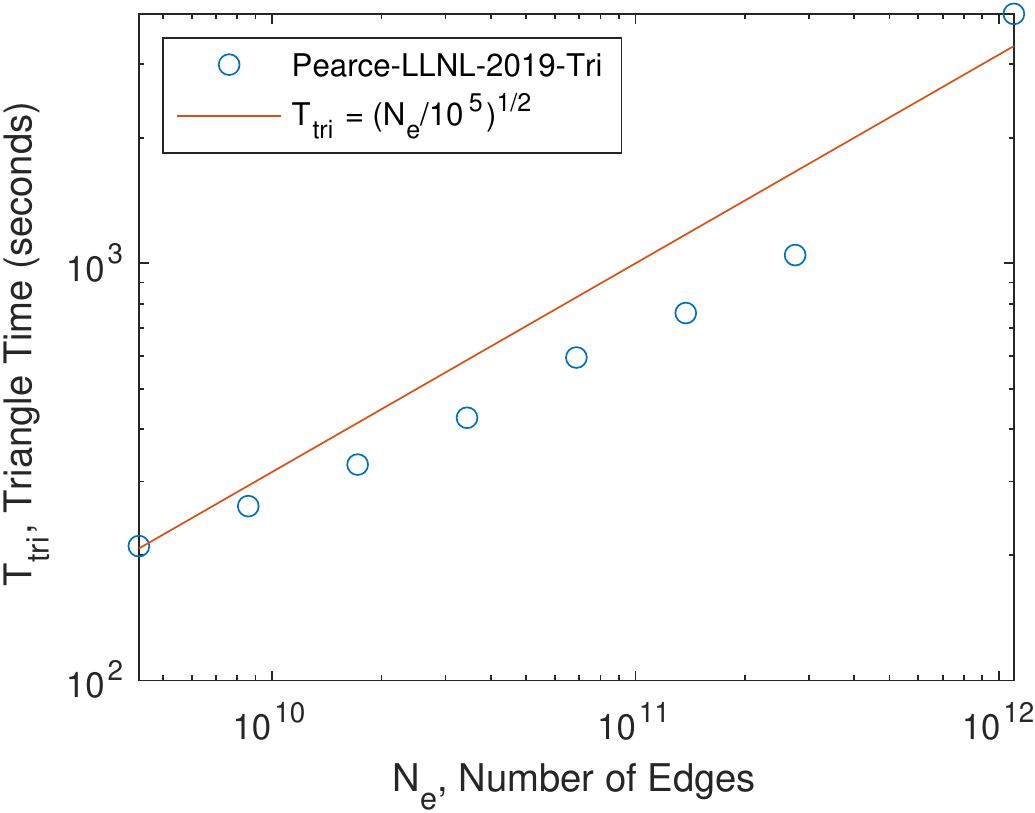}
\includegraphics[width=2.4in]{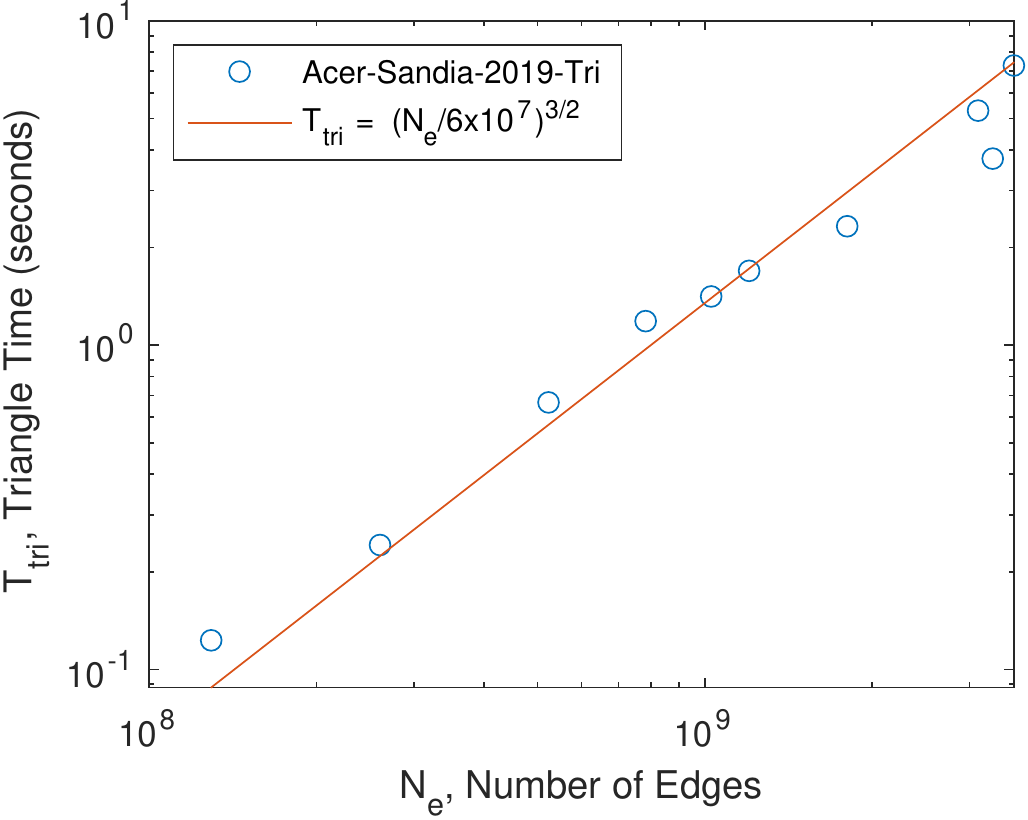}
\includegraphics[width=2.4in]{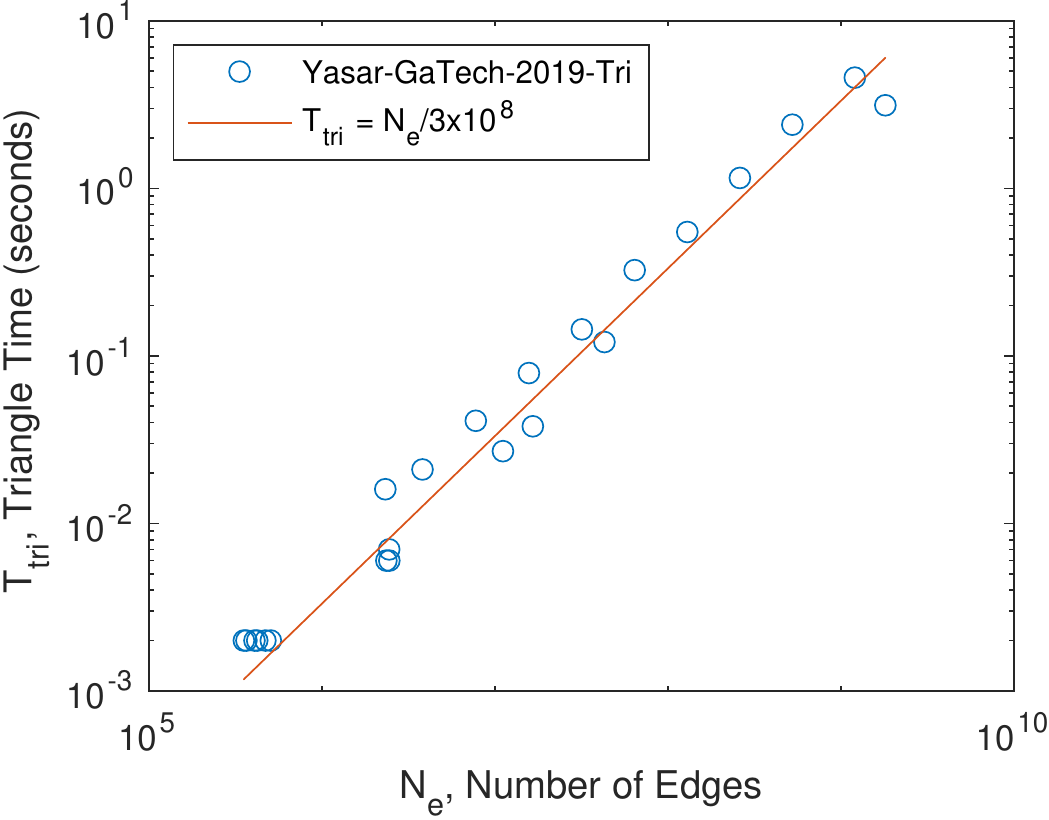}
\caption{Graph Challenge 2019 Champions and Innovation Awards. Triangle counting execution time vs number of edges and corresponding model fits for Pandey-Stevens-2019 \cite{Pandey-Stevens-2019}, Pearce-LLNL-2019 \cite{Pearce-LLNL-2019}, Acer-Sandia-2019 \cite{Acer-Sandia-2019}, and Yasar-GaTech-2019 \cite{Yasar-GaTech-2019}.}
\label{fig:Champions}
\end{figure}

\begin{figure}[ht]
\centering
\includegraphics[width=2.4in]{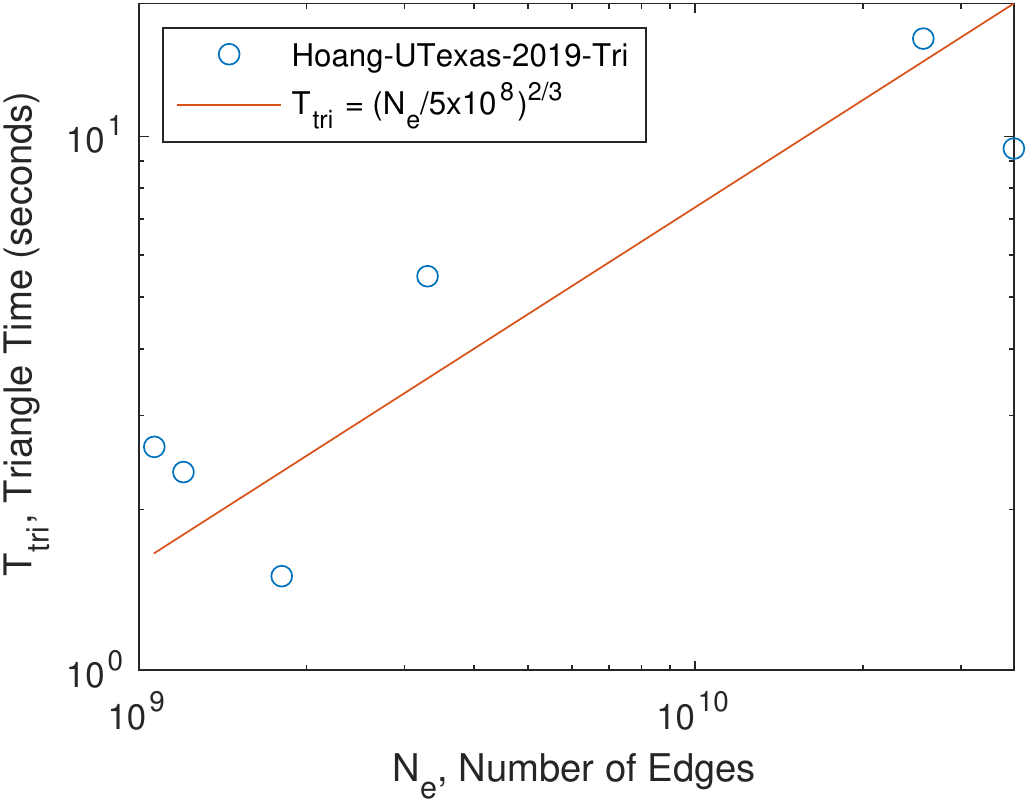}
\includegraphics[width=2.4in]{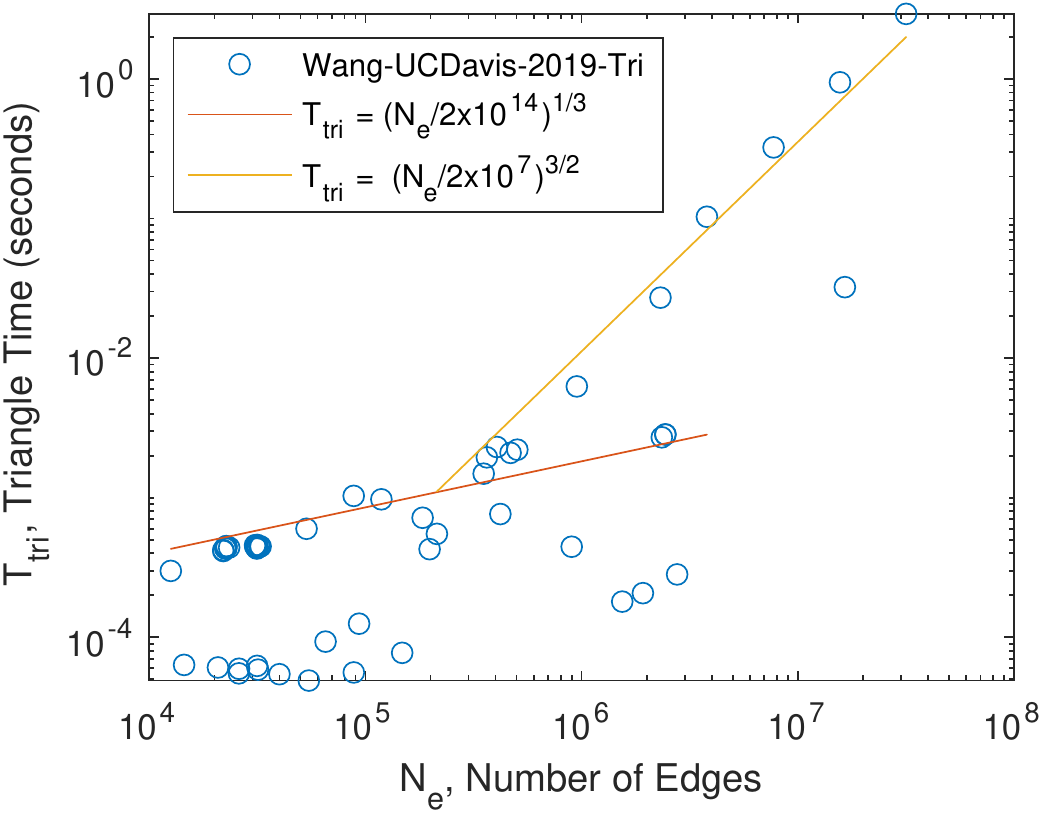}
\includegraphics[width=2.4in]{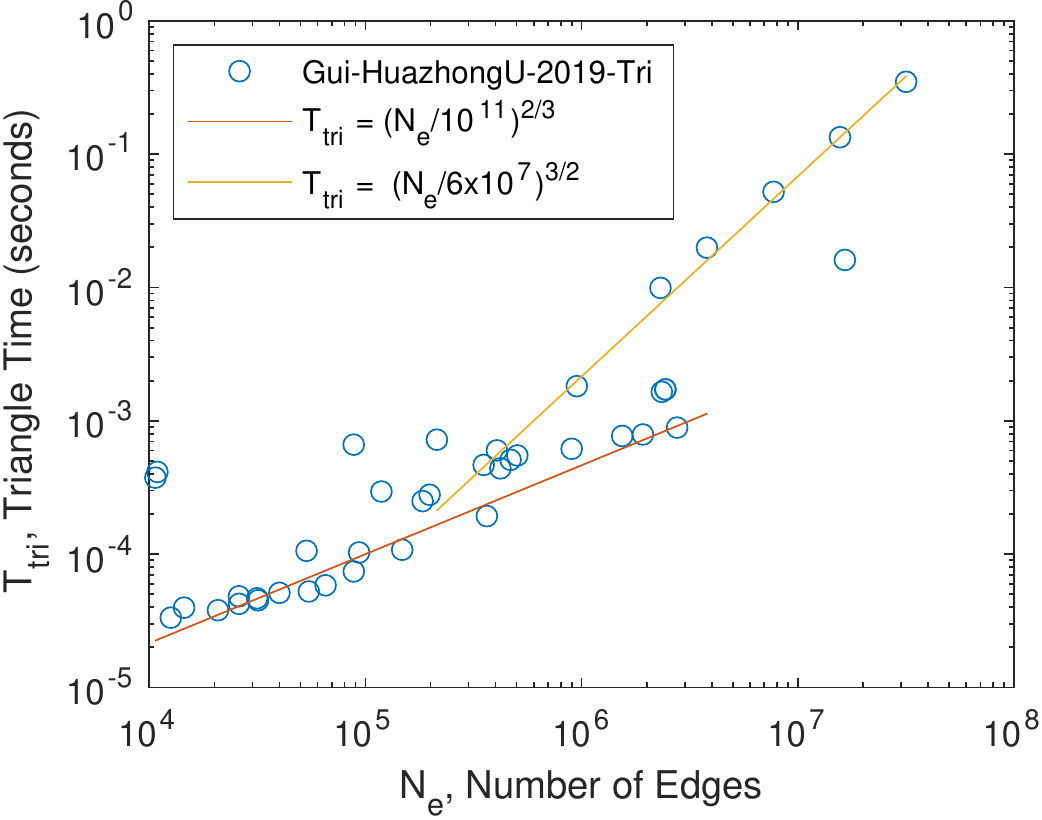}
\includegraphics[width=2.4in]{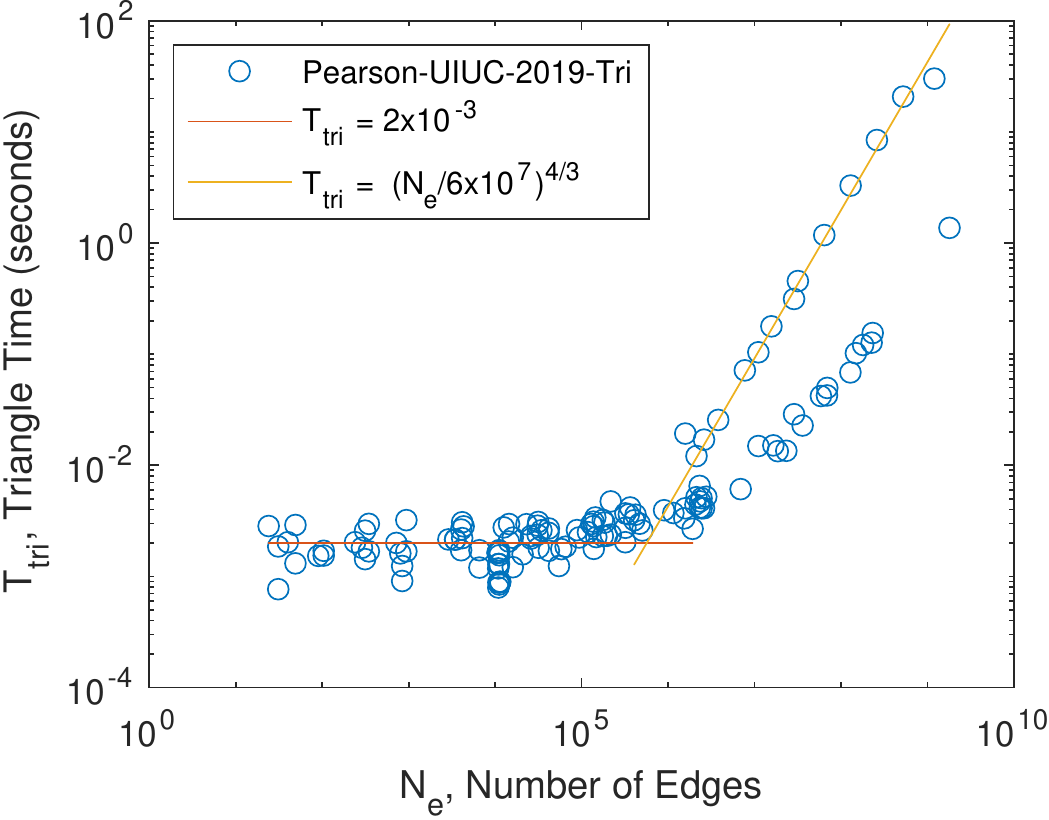}
\caption{Graph Challenge 2019 Student Innovation Award, Finalist, and Honorable Mentions. Triangle counting execution time vs number of edges and corresponding model fits for Hoang-UTexas-2019 \cite{Hoang-UTexas-2019}, Wang-UCDavis-2019 \cite{Wang-UCDavis-2019}, Gui-HuazhongU-2019 \cite{Gui-HuazhongU-2019}, and Pearson-UIUC-2019 \cite{Pearson-UIUC-2019}.}
\label{fig:Champions}
\end{figure}

\end{document}